\documentclass[twocolumn,preprintnumbers,pre,floatfix,superscriptaddress,amsmath,amssymb]{revtex4}
\usepackage{epsfig}
\usepackage{subfigure}
\usepackage{amsmath}
\usepackage{color}
\usepackage{amssymb}
\usepackage{verbatim}
\usepackage{setspace}
\usepackage{graphicx}
\usepackage{dcolumn}
\usepackage{bm}
\usepackage{times}
\input epsf

\begin{document}
\author{Dane Taylor}
	\email{dane.taylor@colorado.edu} 
	\affiliation{Department  of Applied Mathematics, University of Colorado, Boulder, Colorado 80309, USA}
\author{Juan G. Restrepo}
	\affiliation{Department  of Applied Mathematics, University of Colorado, Boulder, Colorado 80309, USA}

\date{\today}

\begin{abstract}
The principal eigenvalue $\lambda$ of a network's adjacency matrix often determines dynamics on the network (e.g., in synchronization and spreading processes) and some of its structural properties (e.g., robustness against failure or attack) and is therefore a good indicator for how ``strongly'' a network is connected.
We study how $\lambda$ is modified by the addition of a module, or community, which has broad applications, ranging from those involving a single modification (e.g., introduction of a drug into a biological process) to those involving repeated additions (e.g., power-grid and transit development). 
We describe how to optimally connect the module to the network to either maximize or minimize the shift in $\lambda$, noting several applications of directing dynamics on networks. 

\end{abstract}

\title{Network connectivity during mergers and growth: optimizing the addition of a module}
\maketitle

\section{Introduction}\label{intro}
Spectral approaches to the analysis of complex networks are becoming increasingly important due to their ability to describe the effect of network structure on dynamical processes. In particular, the principal eigenvalue $\lambda$ of a network's weighted adjacency matrix $A$ ($A_{ij}$ is nonzero if there exists a link from node $i$ to node $j$) is significant for dynamics on networks such as the synchronization of heterogeneous oscillators \cite{Juan_kura, Neg_matrix}, epidemic and information spreading \cite{EigEpidemic}, structural robustness (percolation) \cite{Perc}, the stability of equilibria for certain systems of network-coupled ordinary differential equations \cite{stab_fixed}, the stability of gene expression in genetic networks \cite{gene}, and criticality in network-coupled excitable systems \cite{dans}. 

Given the importance of $\lambda$ in determining the outcome of so many dynamical processes on networks, there has been much interest in modifying $\lambda$ through structural perturbations. In particular, the effect of removing node $j$ can be quantified by its {\it dynamical importance}:
$ I_j =-{\delta\lambda}/{\lambda} \approx{v_ju_j}/{v^Tu}$ \cite{Restrepo},
where $u$ ($v$) is the right (left) eigenvector corresponding to the principal eigenvalue $\lambda$ (i.e., $Au = \lambda u$, $v^TA = \lambda v^T$), and $\delta\lambda$ is the decrease in the principal eigenvalue that would result from the removal of node $j$. As an example application, a node removal strategy targeting nodes with large dynamical importance fragments a network more rapidly than targeting nodes with large degree (number of links) \cite{Restrepo}. Reference~\cite{Milanese} extended these results by finding perturbative expressions for the change in eigenvalue $\delta \lambda$ due to the removal of groups of nodes as well as the addition or deletion of groups of links. Reference \cite{Chauhan} considered a perturbative approach to studying the spectrum of networks with community structure.

In this study, we consider the effect on the largest eigenvalue of a network's adjacency matrix from the addition of a secondary network (referred to as the module or community). As opposed to previous work \cite{Restrepo,Milanese, Chauhan}, we explicitly consider the effect of the module's topology on the resulting eigenvalue and use this information to discuss how one can make optimal connections to either maximize or minimize the effect on $\lambda$. There are many applications where smaller groups adhere to a larger network in social and economical networks \cite{econ} (e.g., the merging of corporations or markets) and biological networks (e.g., modifying a system of biochemical reactions with a drug \cite{drug_target,drug} or the merging of ecosystems \cite{ECO}). For example, recent studies have shown that the effect on the largest eigenvalue of the Jacobian matrix describing interactions in an ecological network due to the addition of a species may be integral to the formation of ecological communities \cite{ECO_add}. 
Moreover, our results offer new insight regarding the prevalence of subgraph motifs (recurrent subgraphs having a frequency higher than expected). While motifs have been cited as essential building blocks in biological networks \cite{motif}, their role is not fully understood. For example, in contrast to several studies indicating that the global stability and robustness of a system is strongly influenced by the structure of motifs \cite{Kaluza}, our study suggests that ``how'' a motif is connected to the remaining network may be as significant as its structure (see Fig. \ref{example2}).

This paper is organized as follows. In Sec.~\ref{sub_add} we describe the problem and introduce variables. In Sec.~\ref{both} we present perturbative approximations for $\delta \lambda$ in terms of spectral properties of $A$ and the module to be added. In Sec.~\ref{tests} we test these approximations on several real networks. In Sec.~\ref{maxi} we discuss how our results can be used to optimize the connections between the original network and module. In Sec.~\ref{disc} we discuss our results, citing several applications of how they may be used to direct dynamics on networks.
These results have application in a range of cases from those in which just a single merger needs to be optimally designed to cases where a large number of small additions need to be optimized to quickly evolve $\lambda$ to a desired value.

\section{module addition}\label{sub_add}
\begin{figure}[t]
\includegraphics[width=0.85\linewidth]{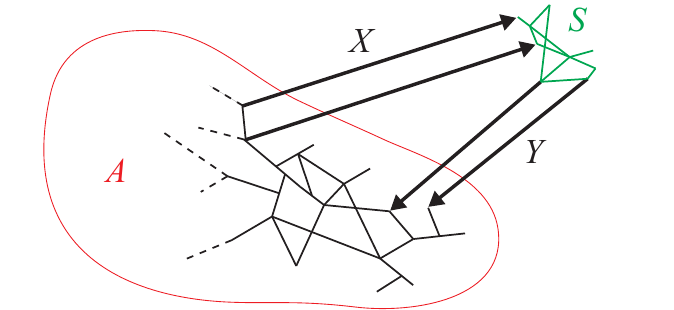} 
\caption{(Color online) A module (described by matrix $S$) is connected to the original network (described by matrix $A$) using directed connections (described by the matrices $X$ and $Y$).}
\label{1}
\end{figure}

We consider the addition of a secondary network, or module, to an existing network, as shown schematically in Fig.~\ref{1}. The original network of size $n$ is described by an $n\times n$ weighted adjacency matrix $A$ such that its entries $A_{ij}$ satisfy $A_{ij}\not=0$ if there exists a link from node $i$ to node $j$ and $A_{ij}=0$ otherwise.  Another network of size $m$ (described by an $m\times m$ adjacency matrix $S$) is to be connected to the original network. We will refer to this secondary network as the {\it module}. In what follows, we will sometimes refer to both the original network and the module by their respective adjacency matrices, $A$ and $S$. 

Assuming that the original eigenvalue problems $Au = \lambda u$ and $v^TA = \lambda v^T$ have been solved, the modified eigenvalue problem after module addition may be formulated as 
\begin{equation}
  \left[  \begin{array}{cc} A & X \\ Y^T & S \end{array} \right] 
\left[ \begin{array}{c}  u +\Delta^{U}\\ \Delta^{L}  \end{array} \right] 
  =  \left( \lambda + \delta \lambda \right)\left[ \begin{array}{c}  u +\Delta^{U}\\ \Delta^{L}  \end{array} \right]  ,
\label{matrix}\end{equation}
where we use the following  definitions: 
(i)  $\delta\lambda\ge0$ denotes the shift in the largest eigenvalue; 
(ii) matrix $X$ ($Y$) is size $n\times m$, has positive entries, and describes all directed links from $A$ to $S$ ($S$ to $A$); 
(iii) $\Delta^{U}$ is a vector of length $n$ which represents the shift in eigenvector $u$;
and (iv) $\Delta^{L}$ is a vector of length $m$ which represents the new eigenvector components.  
For the typical case in which no negative weights are allowed (i.e., $A_{ij}\ge0$), the principal eigenvalues $\{\lambda, \lambda +\delta \lambda\}$ and all entries in $\{u, u+\Delta^{U},\Delta^{L}\}$ are guaranteed to be nonnegative by the Perron-Frobenius theorem for nonnegative matrices \cite{PF}. 
Although in this paper we only consider matrices with positive entries so that the Perron-Frobenius theorem can be applied, in general our analysis only requires that the eigenvalue with largest magnitude $\lambda$ is real and well separated from the remaining eigenvalues. While this is typical for networks with positive links \cite{Chauhan}, it is also observed for networks with negative links provided that they represent a small fraction of the number of links (e.g., see Fig. 6 in \cite{Neg_matrix}).

\subsection{Effect of module addition}\label{both}
We restrict our analysis to cases where the effect of the module addition is small, which will allow us to study its effect as a perturbation to the original eigenvalue problem. This restriction is applicable to describing heavy-sided mergers and applications for which a network is modified gradually, such as the expansion of infrastructure.
Considering the upper and lower blocks of Eq.~(\ref{matrix}) independently and after left-multiplying the top block by the left principal eigenvector $v^T$ (i.e., $v^TA=\lambda v^T$), we obtain
\begin{eqnarray}
\delta \lambda &=& \frac{ v^T X \Delta^{L} }{ v^T (u+\Delta^{U})}\label{dlam1}, \nonumber\\
\Delta^{L} &=& \left[(\lambda+\delta\lambda)I_m-S\right]^{-1}Y^T( u + \Delta^{U}),\label{dl1} \nonumber
\end{eqnarray}
where $I_m$ is the identity matrix of size $m$. 

Assuming that the effect of the module addition is small, we have $\delta\lambda \ll \lambda$, $v^T\Delta^{U} \ll v^Tu$, and $\delta\lambda \ll |\lambda -\lambda_S|$, where $\lambda_S$ is the largest eigenvalue of the module. To first order, we find
\begin{eqnarray}
\delta\lambda  &\approx& \frac{1}{\lambda v^Tu} v^TXK^SY^Tu \label{dlam_inv} , \\
\Delta^{L} &\approx& \lambda^{-1}K^SY^Tu, \label{dil1}
\end{eqnarray}
where we have defined 
$K^S \equiv (I_m-S/\lambda)^{-1}.$
These expressions relate the change in the dominant eigenvalue $\delta\lambda$ to the topology of the added module $S$, the spectral properties of the original networks ($u$, $v$, and $\lambda$), and the way in which the module is coupled to the network by matrices $X$ and $Y$. When the module contains few nodes, approximating $\delta\lambda$ by inverting an $m\times m$ matrix is significantly more efficient and, as we will see, offers more insight than solving the original $(m+n)\times(m+n)$ eigenvalue problem. Using $v=u$ and $X=Y$ for undirected networks, Eq.~(\ref{dlam_inv}) simplifies to $\delta\lambda \approx  \lambda^{-1}(X^Tu)^TK^S(X^Tu)$.  

If the connections between the module and original network are made randomly, we can use Eq.~(\ref{dlam_inv}) to estimate average values of $\delta \lambda$. Suppose that the entries of the matrix $X$ are independent random variables such that $X_{ij} = 1$ with probability $x/(nm)$ and $0$ otherwise,
so that the expected number of links from the original network to the added module is $x$. Similarly, we assume that the entries of $Y$ are independent random variables which are $1$ with probability $y/(nm)$ and $0$ otherwise. By averaging Eq.~(\ref{dlam_inv}) and using the independence of $X$ and $Y$, we find
\begin{align}
\langle \delta \lambda \rangle =  \frac{\bar u\bar v}{ \lambda v^Tu}  \left(\frac{x}{m}\right) \left(\frac{y}{m}\right) \sum_{i,j} K^S_{ij}~, \label{average}
\end{align}
where $\bar u = n^{-1}\sum_{j=1}^n u_j$ and $\bar v = n^{-1}\sum_{j=1}^n v_j$. Thus, in addition to properties dependent on the original network, $\langle \delta \lambda \rangle$ is proportional to the product of the relative number of connections to and from the module ($xy/m^2$) and on the sum of elements in the matrix $K^S$. Moreover, for large $\lambda/\lambda_S$ we have $\sum_{ij} K^S_{ij} \approx m$, the number of nodes in $S$. While this expression provides us with the average $\langle\delta \lambda\rangle$ when $X$ and $Y$ are chosen randomly, as discussed in Sect. \ref{maxi}, strategically selecting connection matrices ($X$,$Y$) (e.g., to maximize $\delta \lambda$) can lead to significant variations in $\delta\lambda$ for a given module.

For the optimization objectives explored later in this text, as well as situations in which computing $K^S$ is inconvenient, it is useful to represent Eqs.~(\ref{dlam_inv}) and (\ref{dil1}) using a series expansion for $K^S$. For $\lambda_S < \lambda$, we have $K^S = (I_m-S/\lambda)^{-1}\approx\sum_{j=0}^k (S/\lambda)^j$.  We thus define the $k$-th order approximations as
\begin{eqnarray}
\delta \lambda_k &=& \frac{1}{ \lambda v^Tu} \sum_{j=0}^{k-1}\lambda^{-j}{ v^T XS^jY^Tu}  , \label{dlam_dir}\\
\Delta^{L}_k  &=& \sum_{j=0}^{k-1} { \lambda^{-(j+1)}S^jY^Tu} \label{dLk} .
\end{eqnarray}
Because the matrices ($X$,$Y$) are often sparse and the module is often much smaller than the network, Eq.~(\ref{dlam_dir}) is typically very computationally efficient. We note that for large enough $k$, the error introduced by Eq.~(\ref{dlam_inv}) dominates the error of series truncation in Eqs.~(\ref{dlam_dir}) and (\ref{dLk}). No gain was found by using $k>4$ in the experiments that are to follow.

\begin{table}[t]
\begin{center}
\vspace{0.2cm}
\begin{tabular}{|c|c|c|c|c|}
\hline
Network and reference&  N   & $\langle d\rangle$ &  $\lambda$ & $\lambda_2$ \\
\hline
Neural network of C. elegans  \cite{cElegans} & $297$ & $7.9$ &$9.2$ & $5.7$  \\
\hline
Network of political blogs \cite{poliblog} &  $1490$& $12.8$ & $34.4$&$26.8$\\
\hline
Yeast PPI network \cite{Yeast} &  $2361$ & $5.6$ & $12.1$&$9.4$\\
\hline
Word association network \cite{Word} & $5018$ & $12.7$& $13.4$ &$10.2$ \\
\hline
\end{tabular}
\caption{Test networks used and their characteristics: number of nodes $N$; mean degree $\langle d \rangle$; largest eigenvalue $\lambda$; and second largest eigenvalue $\lambda_2$.}
\label{networksused}
\end{center}
\end{table}

\subsection{Numerical tests on real networks}\label{tests}
\begin{figure}[t]
\begin{center} 
\includegraphics[width=.95\linewidth]{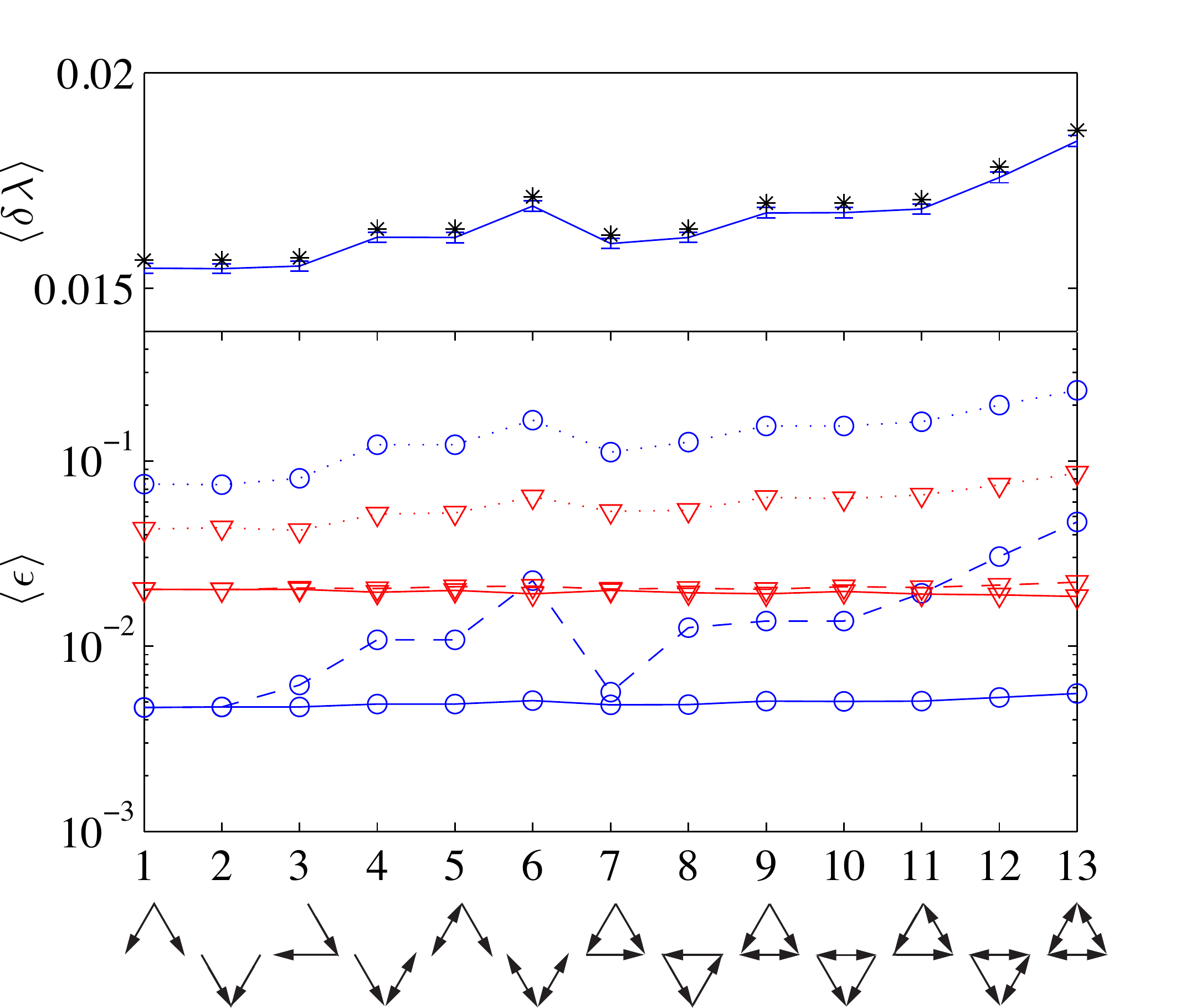} 
\end{center}
\caption{(Color online) $\delta\lambda$ and approximation errors, $\epsilon$, were averaged over $10^4$ realizations of connecting a three-node module to the networks in Table \ref{networksused} using 10 random links (see text). Equation (\ref{average}) (stars) is shown to be accurate in the upper panel for typical results for the neural network of C. elegans. The average relative error $\langle\epsilon\rangle$ for the neural network of C. elegans (circles) and a network of political blogs (triangles) are given in the lower plot, where Eq.~(\ref{dlam_inv}) (sold lines) and Eq.~(\ref{dlam_dir}) with $k = 1$ (dotted) and $k = 2$ (dashed) are shown. }
\label{motif}
\end{figure}

We test our approximations by considering module additions to four networks: a neural network of C. elegans \cite{cElegans}; a network of political blogs \cite{poliblog}; a network of protein-protein interactions in the organism S. cerevisiae (i.e., brewers/bakers yeast) \cite{Yeast}; and a network of associations between words \cite{Word}. Their characteristics are summarized in Table~\ref{networksused}.
We begin by examining the average effects for adding a module using random connections. First, matrices were constructed by randomly selecting $10$ entries in $X$ and $Y$ to be $1$ and the rest to be $0$ (i.e., in the previous notation, $x=y=10$). For each realization $\delta\lambda^{\mbox{act}}$, the actual eigenvalue shift [i.e. solving Eq.~(\ref{matrix})], was compared to our approximations given by Eqs.~(\ref{dlam_inv}) and (\ref{dlam_dir}). In the top panel of Fig.~\ref{motif}, Eq.~(\ref{average}) (stars) is shown to accurately predict the numerically-observed average $\langle\delta\lambda\rangle$ (solid line) for connecting the modules to the directed neural network of C. elegans using $10^4$ realizations of ($X$,$Y$). Average values for the relative error $\epsilon=(\delta\lambda-\delta\lambda^{\mbox{act}})/\delta\lambda^{\mbox{act}}$ are plotted in the bottom panel for both the neural network (circles) and an network of political blogs (triangles) (see Table\ref{networksused}) for all 13 non-isomorphic, directed modules of size 3. (Results for the other networks were found to be similar and are omitted for clarity.) Solid lines correspond to Eq.~(\ref{dlam_inv}), while dotted (dashed) lines correspond to Eq.~(\ref{dlam_dir}) with $k = 1$ ($k = 2$).
 
It can be observed in Fig.~\ref{motif} that $\langle\delta\lambda\rangle$ changes substantially for the different three-node modules (for all networks, $\langle\delta\lambda\rangle$ typically increased $\sim20\%$ from module 1 to module 13). Observe that the average error $\langle\epsilon\rangle$ of Eq.~(\ref{dlam_dir}) when the module structure is not used [$k=1$ (dotted lines in lower plot)] is strongly correlated with $\langle\delta\lambda\rangle$ (upper plot). This is to be expected as the error from neglecting module structure should be related to that structure's ability to modify $\lambda$. Note that for the political blog network (triangles), using $k=2$ in Eq.~(\ref{dlam_dir}) is nearly as accurate as directly using Eq.~(\ref{dlam_inv}). As previously mentioned, for large enough $k$ the dominant source of error comes from neglecting higher orders of $\delta\lambda/\lambda$ in the derivation of Eq.~(\ref{dlam_inv}) [as opposed to series truncation in Eq.~(\ref{dlam_dir})]. 

\begin{figure}[t]
\begin{center}
\includegraphics[width=\linewidth]{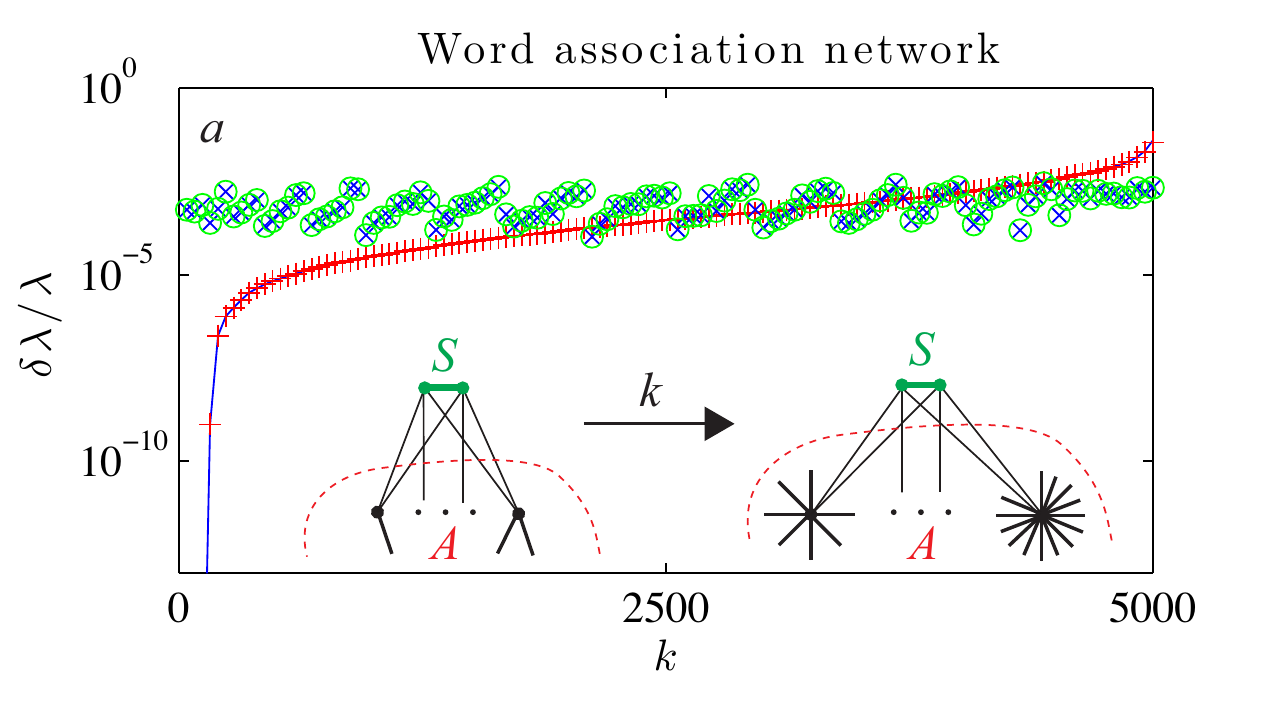}\\
\includegraphics[width=\linewidth]{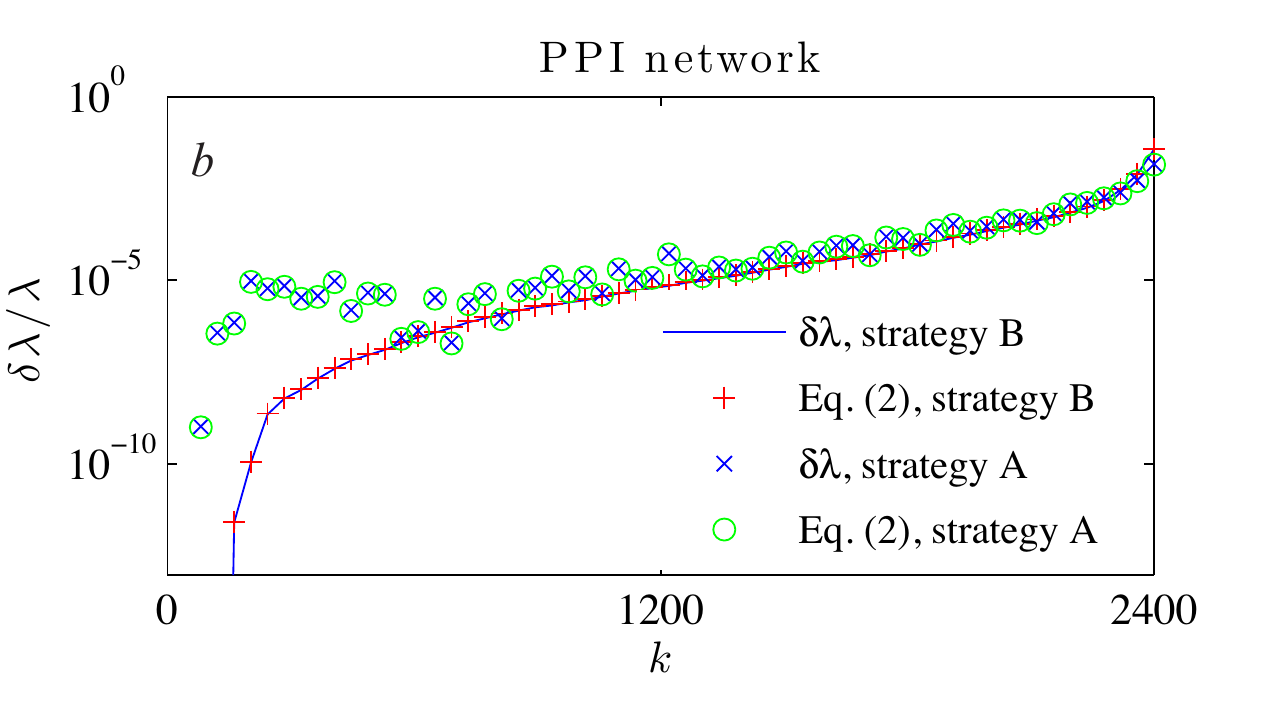}  
\end{center}
\caption{(Color online) Eigenvalue shift $\delta \lambda$ for connecting a two-node module to (a) the word association network and (b) the PPI network. Equation (\ref{dlam_inv}) (crosses) agrees well with actual values ${\delta\lambda}^{\mbox{act}}$ (solid line) for strategy B. The x's and circles show the same respective quantities for strategy A. As indicated by the drawing, increasing $k$ corresponds to connecting the module to nodes with increasing degrees (strategy A) or eigenvector entries (strategy B).}
\label{2strategies}
\end{figure}

The validity of our approximations for specific connections is shown by considering the addition of two bidirectionally-linked nodes ($m=2$) to an undirected protein-protein interaction (PPI) network and a directed network of word associations (see Table~\ref{networksused}). In order to illustrate the dependence of $\delta\lambda$ on the matrices $X$ and $Y$, we will consider two connection strategies: connecting the module to nodes with either (A) increasing nodal degrees or (B) increasing eigenvector entries. For strategy A, the nodes in the original network are ordered so that the in-degrees monotonically increase: $d_1^{in}\leq d_2^{in} \leq \dots \leq d_N^{in}$. Then for  $k\in\{1,2,...,N-20\}$, we establish a directed link from nodes $\{k, k+1,\dots,k+20\}$ to both nodes in the module. The nodal out-degrees are then ordered such that $d_{i_1}^{out}\leq d_{i_2}^{out} \leq \dots \leq d_{i_N}^{out}$, and links are made to nodes $\{i_k, i_{k+1},\dots,i_{k+20}\}$ from both nodes in the module. The case $k=0$ corresponds to connecting the network nodes with smallest $d^{in}$ to both module nodes, which in turn connect to the nodes with smallest $d^{out}$; whereas the case $k = N-20$ corresponds to connecting the nodes with largest $d^{in}$ to both module nodes, both of which in turn connect to the nodes with largest $d^{out}$ (shown schematically in Fig.~\ref{2strategies}a). 

For strategy B, we now order the nodes in the original network in order of increasing entries of the left eigenvector $v$ so that $v_1\leq v_2 \leq \dots \leq v_N$. As before, for $k\in\{1,2,...,N-20\}$, we connect nodes $\{k, k+1,\dots,k+20\}$ in the network to both module nodes, both of which in turn connect to nodes $\{i_k, i_{k+1},\dots,i_{k+20}\}$, where the indices $i_j$ now correspond to the ordering of the right eigenvector entries such that $u_{i_1}\leq u_{i_2} \leq \dots \leq u_{i_N}$. For both strategies, the indices simplify for undirected networks, for which we have $u=v$, $d^{out}=d^{in}$, and $i_k=k$.

In Fig.~\ref{2strategies}, $\delta \lambda$ is plotted for strategies A and B as a function of the parameter $k$ for both (a) the directed word-association network and (b) the undirected PPI network. For strategy B, the crosses show the approximation given by Eq.~(\ref{dlam_inv}) and the solid line shows the numerically-calculated value from directly solving the eigenvalue problem Eq.~(\ref{matrix}). The x's and circles respectively show the same quantities for strategy A. The first observation is that the approximation for $\delta\lambda$ works well, with only a small deviation as the perturbation becomes large (not shown).  One can observe that strategy B is superior for yielding either large or small $\delta \lambda$ for both networks. However, the two strategies are similar for producing large $\delta\lambda$ for the PPI network in Fig.~\ref{2strategies}b. This is expected when the first-order approximations to the eigenvectors ($u_i\propto d_i^{out}$ and $v_i\propto d_i^{in}$ \cite{Restrepo}) are valid. The results of this experiment suggest that it may be useful to devise connection strategies to systematically maximize (or minimize) $\delta \lambda$.

\section{Optimizing connections}\label{maxi}  
The issue of efficiently decreasing $\lambda$ by removing nodes or links from a network has been recently addressed \cite{Restrepo}, where it was found that when removing a single node, $\lambda$ is most decreased by removing the node with largest dynamical importance. We consider a closely related issue: given a module $S$ to be added to a network $A$ with given constraints (such as a fixed number of connections), how should the links between the network and module be chosen to either maximize or minimize $\delta\lambda$? Given some set of constraints and staying within our previous assumptions, we will look for matrices ($X$,$Y$) that maximize (or minimize) $\delta \lambda$ in Eq.~(\ref{dlam_inv}). In the examples that follow, it is helpful to assume that the node indices are now ordered such that the left eigenvector entries are in decreasing order: $v_1 \ge v_2 \ge \dots \ge v_n\ge0$. In addition, the entries of the right eigenvector are ordered using indices $\{l_i\}$ so that $u_{l_1} \geq u_{l_2} \geq \dots \geq u_{l_n}\ge0$. (If $A$ is symmetric, $u = v$ and $l_i = i$.) We will present our optimization methodology for two examples, yet the techniques presented are general and have potential application beyond these particular constraints.

\subsection{Example I: multiple links per module node}
 In the first example we assume that the number of connections from the original network to the module, $x$, and the number of connections from the module to the original network, $y$, are fixed and less than $n$, the number of nodes in the original network. It is also assumed that all links have strength one (i.e., $X_{ij},Y_{ij}\in\{0,1\}$) and multiple links per module node are allowed. 

\begin{figure}[h]
\begin{center}
\includegraphics[width=.5\linewidth]{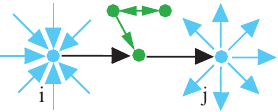}  
\end{center}
\caption{(Color online) A typical optimal connection for example (I): node $i$ (a {\it point of contraction} with large left eigenvector entry $v_i$) points to a node in the module, which in turn points to node $j$ (a {\it point of expansion} with large right eigenvector entry $u_j$). }
\label{example1}
\end{figure}

The right hand side of Eq.~(\ref{dlam_inv}), which approximates the quantity to be maximized, is proportional to $\sum_{i,j} (X^T v)^T_i K^S_{ij} (Y^T u)_j$. This sum can be maximized by (i) finding indices $a$ and $b$ such that $K_{ab} = \max_{ij}\{ K_{ij}\}$ and (ii) choosing $X$ and $Y$ to make $(X^T v)^T_a$ and $(Y^T u)_b$ as large as possible. The scalar $(X^T v)^T_a$ is maximized by placing the $x$ ones in the $a$-th column of $X$ and in positions $1,2,\dots, x$ corresponding to the largest values of $v$, while $(Y^T u)^T_b$ is maximized by placing the $y$ ones in the $b$-column of $Y$ and in positions $l_1, l_2,\dots, l_y$ corresponding to the largest values of $u$. In this way, $(X^T v)^T_i = \delta_{ia} \sum_{j=1}^x v_{j}$ and $(Y^T u)^T_i = \delta_{ib} \sum_{j=1}^y u_{l_j}$, where $\delta_{ij}$ is Kronecker's delta. The maximum of Eq.~(\ref{dlam_inv}) is then 
\begin{equation}\label{max}
\delta \lambda^{max} \approx \frac{K^s_{ab}}{\lambda v^Tu} \sum_{i=1}^y v_i \sum_{j=1}^x  u_{l_j}
\end{equation}
This result implies $\delta \lambda$ may be maximized for the constraints of example (I) by connecting the $x$ nodes with the largest left eigenvector entries $v_i$ in the original network to a single node in the module (having index $a$), and by also originating all links from the module to the original network from a single module node (having index $b$) to the $y$ nodes in the original network with the largest entries of the right eigenvector $u$. For large values of $\lambda/\lambda_s $, the maximum entry of matrix $K^s$ is typically in its diagonal, yielding $a=b$ as shown in Fig.\ref{example1}. 

For a heuristic interpretation of this result, let us assume that $A_{ij}\in\{0,1\}$ and denote $L_i^{ {o},p}=\sum_{j} (A^p)_{ij}$ and $L_i^{ {t},p}=\sum_{j} (A^p)_{ji}$ as the number of paths of length $p$ originating from and terminating at node $i$, respectively. Thus $L_p=\sum_{ij} (A^p)_{ij}$ is the total number of paths of length $p$. These quantities satisfy $||L_i^{ {o},p}||_2^{-1}L_i^{ {o},p}\to u_i$, $||L_i^{ {t},p}||_2^{-1}L_i^{ {t},p}\to v_i$, and $L_{p+1}/L_p \to \lambda$ as $p \to \infty$ \cite{Golub}. Therefore connecting nodes with large $v_i$ (which receive many paths) to nodes with large $u_i$ (which distribute many paths) will have the largest impact in how $L_p$ grows with $p$, which determines $\lambda$. We therefore define a node $i$ with large $v_i$ as a {\it point of contraction} and a node $j$ with large $u_j$ as a {\it point of expansion}. 
Therefore our result for example (I) is that the effect of the whole module is to act as a bridge from points of contraction to points of expansion in the original network. 

\subsection{Example II: one link per node}
\begin{figure}[b]
\begin{center}
\includegraphics[width=.95\linewidth]{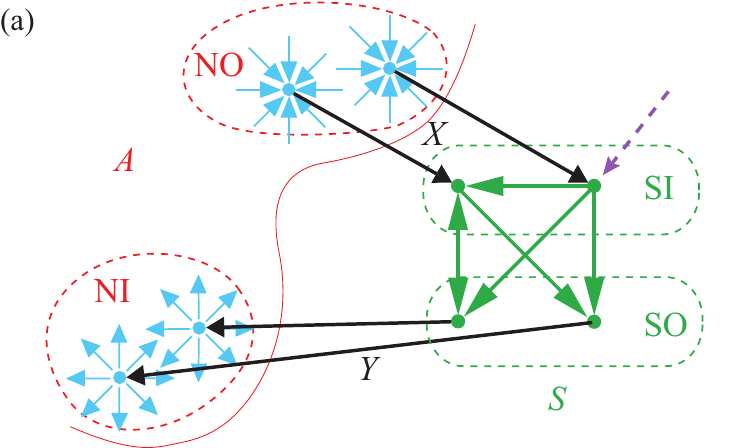}\\
\includegraphics[width=\linewidth]{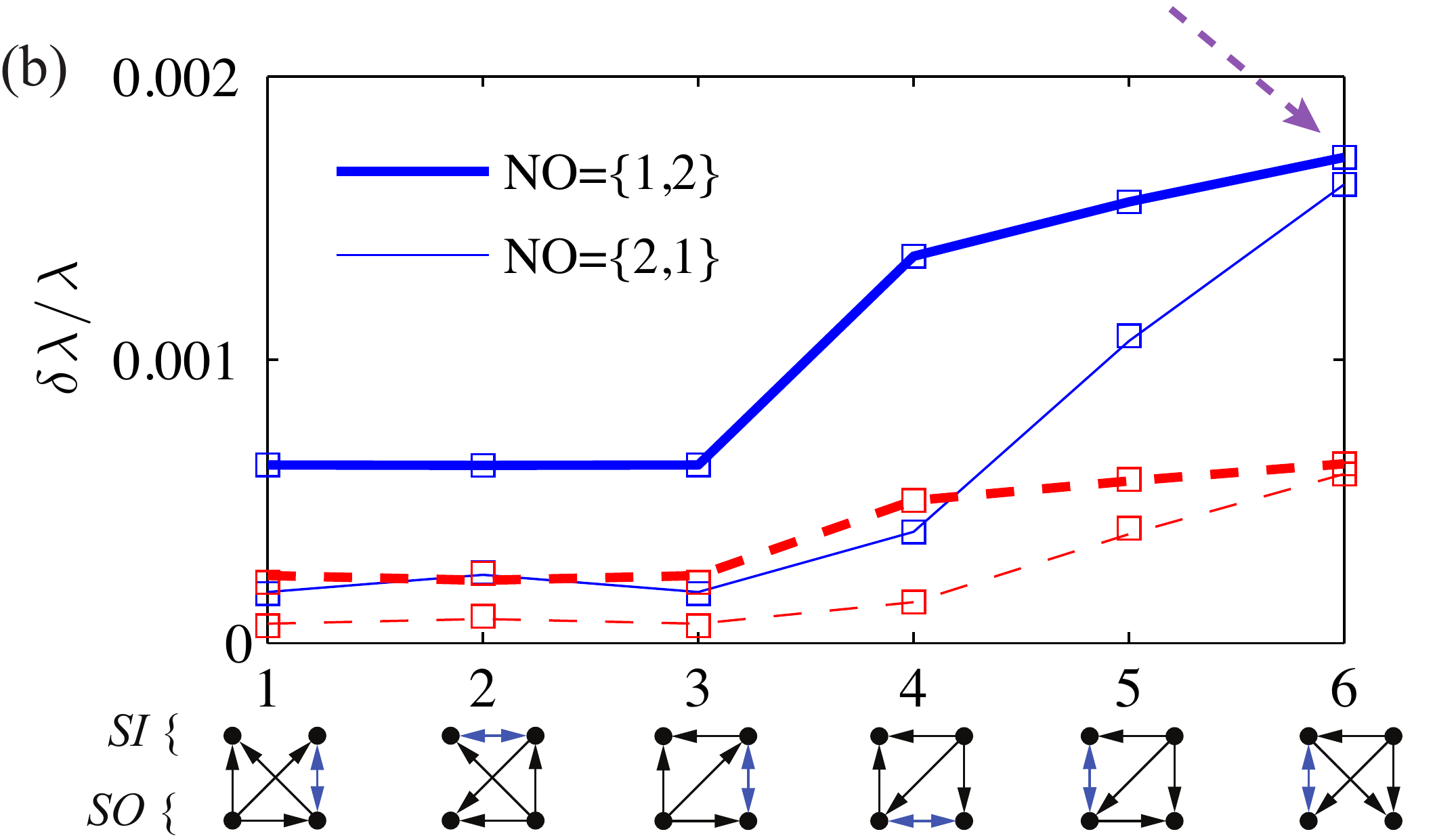} 
\end{center}
\caption{(Color online) (a) Typical optimal link selections for example (II) for $x=y=2$. Two points of contraction (NO) link to two module nodes (SI) and the remaining two module nodes (SO) link to two points of expansion (NI). The module also contains a directed complete bipartite graph pointing from SI to SO.
(b) Under the restrictions of example (II), the module in Fig.~\ref{example2}a was connected to the neural network for C. elegans using various orientations. Solid lines indicate letting $\mbox{NI}=\{l_1,l_2\}$ and either $\mbox{NO}=\{1,2\}$ (thick) or $\mbox{NO}=\{2,1\}$ (thin). Symbols show Eq.~(\ref{dlam1}). Approximating points of contraction (expansion) by nodes with large $d^{in}$ ($d^{out}$) also offers a decent strategy (dashed).}
\label{example2} 
\end{figure}

In the second example we require that, in addition to a fixed number of links $x$ and $y$ with unity strength, no more than one link can be added to a particular node in the module or original network. Because undirected links are equivalent to two links and violate our constraint, it is reasonable (although not necessary) to assume that the network and module are directed. To treat this case, we maximize successive terms in Eq.~(\ref{dlam_dir}). The first term, $v^T X Y^Tu/\lambda$, vanishes since any entry of $XY^T$ is nonzero only if there is a module node that has links both to and from the network, a situation which is not allowed by our constraint. Therefore, we maximize the next term, $(X^T v)^T S (Y^Tu)/\lambda^2$. As shown in Fig.~\ref{example2}a, let us denote the set of nodes in the original network that point to the module as NO (network outgoing), the set of nodes in the module that are pointed to by the original network as SI (module incoming), the set of nodes in the module that point to the original network as SO (module outgoing), and the set of nodes in the original network that are pointed to by the module as NI (network incoming). Because no node can have more than one new link, there is a one-to-one correspondence between nodes in NO and nodes in SI. The index of nodes in SI will be represented as $i_j$, where node $j$ in NO points to node $i_j$ in SI. We have
$(X^T v)^T_{i_j} = v_{j}$ if $j \in$ NO and $i_j \in$ SI, and $0$ otherwise. With a similar notation for SO and NI, we have $(Y^Tu)_{m_k} = u_k$ if $m_k \in$ SO and $k \in$ NI, and $0$ otherwise. It follows that Eq.~(\ref{dlam_dir}) yields 
\begin{equation}
\delta \lambda_2 = \frac{1}{\lambda^{2}v^Tu}  \sum_{j \in \mbox{NO}}v_j \sum_{k \in \mbox{NI}} S_{i_jm_k} u_k. \nonumber
\end{equation}
This expression is maximized if $S$ contains a directed complete bipartite graph for disjoint subsets $SI$ and $SO$ such that every node in SI points to every node in SO (see Fig.~\ref{example2}a).  Assuming that one can be found, we may set $S_{i_jm_k}= 1$ and look for sets NO and NI that solve
\begin{equation}\label{MAX}
\delta \lambda_2^{max} =\left(\frac{1}{\lambda^{2}v^Tu}  \right)\max_{ \mbox{NO}\cap \mbox{NI} = \emptyset}\left(\sum_{j\in \mbox{NO}}v_j\sum_{k\in \mbox{NI}}u_k\right). 
\end{equation}
Let $Q=\{i\}_{i=1}^x\cap\{l_i\}_{i=1}^y$. If $Q=\emptyset$ then Eq.~(\ref{MAX}) is solved by letting $\mbox{NO}=\{i\}_{i=1}^x$ and $\mbox{NI}=\{l_i\}_{i=1}^y$, which yields
\begin{equation}\label{max4}
\delta \lambda^{max} \approx  \frac{1}{\lambda^2 v^Tu} \sum_{j=1}^x v_{j}\sum_{i=1}^y u_{l_i}.
\end{equation}
As indicated in Fig.~\ref{example2}a, this corresponds to selecting nodes of contraction for NO and nodes of expansion for NI. 

The significance of link choices for maximizing $\delta\lambda$ is shown in Fig.~\ref{example2}b, where the module in Fig.~\ref{example2}a was added to the neural network of C. elegans with constant node selections for NI and NO but using several module orientations (defined as a particular choice for the disjoint sets SO and SI in the module, and shown in the horizontal axis of Fig.~\ref{example2}b). The solid lines show $\delta\lambda/\lambda$ found numerically using $\mbox{NI}=\{l_1,l_2\}$ and either $\mbox{N0}=\{1,2\}$ (thick) or $\mbox{N0}=\{2,1\}$ (thin) (see next paragraph for discussion). Symbols indicate $\delta\lambda/\lambda$ found using Eq.~(\ref{dlam1}). One can observe that our maximization strategy for example (II) (Fig.~\ref{example2}a) does in fact maximize $\delta\lambda$ (see  orientation 6). An important practical issue is that the eigenvectors may be unknown and require estimation using local information. One can observe that attempting to maximize $\delta\lambda$ using the first-order approximations $v_i\varpropto d_i^{in}$ and $u_i\varpropto d_i^{out}$ \cite{Restrepo} may also be a good strategy (dashed lines). If necessary, a more refined approximation for the eigenvectors may be sought (e.g., using second-order neighbors \cite{Milanese}).

It is important to note that we have so far neglected higher order terms of Eq.~(\ref{dlam_dir}) in addressing example (II), which are responsible for the difference in $\delta\lambda$ for the permutation $\mbox{NO}=\{1,2\}$ or $\mbox{NO}=\{2,1\}$. Attempting to maximize the third term of the series in Eq.~(\ref{dlam_dir}) (which is proportional to $v^TXS^2Y^Tu$) while using the nodes of contraction, 1 and 2 (with $v_1\approx0.58$ and $v_2\approx0.23$), we see that the more-dominant point of contraction (node 1) should link to the module node indicated by the dashed arrow in Fig.~\ref{example2}a. (Note that there is a path of length 2 stemming from this node to each node in SO, whereas there are none for the other node in SI.) Unlike permuting nodes in SI, permuting nodes in SO had little effect for this network since $u_{l_1}\approx u_{l_2}\approx0.23$.

Up to this point we have assumed $Q=\emptyset$, where $Q$ is defined just after Eq.~(\ref{MAX}); however this may not always be the case. For example, as more links are made (i.e., for increasing $x,y$), one would expect some nodes to have large values for both $v_i$ and $u_i$. This may also occur for networks with correlations between $d^{in}$ and $d^{out}$ and, in fact, always occurs for undirected networks where $l_i=i~\forall~i$. For these situations, nodes in $ Q$ must be allocated to either NO or NI and additional nodes must be selected. Considering the limiting case of an undirected network under the constraints of example (II), maximization of the second-order term in Eq.~(\ref{dlam_dir}) indicates that we should choose $\mbox{NO,NI}\subset\{i\}_{i=1}^{x+y}$. (Recall that the first-order term is zero by our constraints.)  The allocation of these indices should then correspond to successively maximizing the third-, fourth-, ..., $k$th-order terms until all degrees of freedom have been exhausted. 
While this strategy of successive maximization does not guarantee the optimal connections (which would require considering all possible links between $S$ and $A$), it is computationally efficient and ensures a near-optimal solution.

\section{Discussion}\label{disc}

While we have presented an efficient strategy for maximizing $\delta\lambda$ for the addition of a module under two examples of constraints, our methodology is general and is thus applicable for many constraints not discussed here.
For example, the problem of minimizing $\delta\lambda$ under the constraints of example (II) may be solved by minimizing successive terms of Eq.~(\ref{dlam_dir}). Heuristically, this corresponds to connecting nodes in $A$ with small values of $v_n$ to the module, and then from the module to nodes in $A$ with small values of $u_n$. The module should also be oriented so as few links as possible point from SI to SO. We now discuss several applications of using module addition(s) to direct dynamics on networks.

Increasing $\lambda$ has many real-world applications. For example, because $\lambda$ relates to the ability of network-coupled oscillatory systems to synchronize \cite{Juan_kura, Neg_matrix}, one or several module additions to increase $\lambda$ may be useful to promote synchronization in, for example, a biological process or power grid. 
Moreover, epidemic thresholds of spreading processes on networks are often dependent on $\lambda^{-1}$ \cite{EigEpidemic}. Increasing $\lambda$ can increase the connectivity of a network, improving flow and reducing the epidemic threshold. This may be useful, for example, if one wants to improve communication over a social network or routing-system. The related problem of percolation on networks (where nodes and/or links are randomly removed) is also related to $\lambda^{-1}$ \cite{Perc}. Increasing $\lambda$ can increase a network's robustness against network degradation under failure, blackout, jamming, or attack. 

For other dynamical systems, it is beneficial to have a small value for $\lambda$. For example, the instability of equilibria for a system of network-coupled ordinary differential equations (ODEs) (e.g., interactions in a metabolic network) is related to the largest eigenvalue of a weighted adjacency matrix defined in terms of the system's Jacobian \cite{stab_fixed} (i.e., if $\lambda<1$, then the equilibria are stable). When the eigenvalue of the Jacobian matrix with largest magnitude is real and well separated from the bulk of the spectrum, our method is applicable. For example, besides choosing appropriate link weights to keep $\delta\lambda$ small, choosing optimal connections and module orientation (as shown in Sect.~\ref{maxi}) may also aid in preserving the stability of equilibria for a system undergoing modification. Such analysis may be relevant, for example, in understanding the formation of ecological communities for which the largest eigenvalue of the system's Jacobian has already been suggested to guide the network's evolution under species additions and subtractions \cite{ECO_add}. 

Future applications of our results are also not limited to network dynamics for which the dependency on $\lambda$ is currently well-established. For example, minimizing $\delta\lambda$ for a module addition may present an effective strategy for minimizing global effects during a network modification. Possible applications may include aiding the in development of systems-level drug design by indicating candidate drug targets that are less invasive (e.g., nodes with middle-valued degrees are typical \cite{drug_target}). The importance of developing mathematical approaches for this promising field are often mentioned \cite{drug}. Another open question is the implications of our results on the prevalence of subgraph motifs, which have been proposed to be the basic building blocks of biological networks \cite{motif}. In contrast to several studies showing that global dynamics of a system can depend on the structure of subgraph motifs \cite{Kaluza}, our results suggest that ``how'' motifs are connected in the network may be as important as their structure. 

The work of D. T. and J. G. R. was supported by NSF Grant No. DMS-0908221. 

\bibliographystyle{plain}

\end{document}